\documentclass{PoS}
\usepackage{amsmath}

\title{Experimental studies at low $Q^2$ of the spin structure of the nucleon at Jefferson Lab}

\ShortTitle{JLab low $Q^2$ spin structure}

\author{\speaker{A. Deur}\thanks{This material is based upon work supported by
the U.S. Department of Energy, Office of Science, Office of Nuclear
Physics under contract DE--AC05--06OR23177}\\
        Thomas Jefferson National Accelerator Facility\\
        E-mail: \email{deurpam@jlab.org}}

\abstract{
We summarize the experimental program of Jefferson Lab that studies the nucleon spin structure at low $Q^2$.
This program completes the precise experimental mapping of the nucleon spin structure functions $g_1(\nu, Q^2)$ and 
$g_2(\nu, Q^2)$  and their moments started at SLAC, CERN and DESY at high $Q^2$, and continued at Jefferson Lab 
at intermediate $Q^2$.
The results presented cover the domain where Chiral Effective Field Theory ($\chi$EFT) should describe the 
strong interaction. They provide a comprehensive set of benchmark measurements for $\chi$EFT. 
The preliminary conclusion is that nucleon spin structure data are still challenging for  $\chi$EFT in spite of 
the notable improvements in these calculations. 
}

\FullConference{The 9th International workshop on Chiral Dynamics\\
		17-21 September 2018\\
		Durham, NC, USA}

\begin{document}

\section{Nucleon spin structure studies at Jefferson Lab}

Jefferson Lab (JLab), is an accelerator situated in Virginia, USA, that produces an up-to 12 GeV 
electron beam serving four experimental halls (A, B, C and D). Its main purpose is to
study Quantum Chromodynamics (QCD) using high-energy electrons scattering off fixed targets. 
Prior to the start of the 12 GeV program in 2014, JLab provided a 6 GeV beam to three experimental halls  
(A, B and C). The low-$Q^2$ spin structure experimental program at JLab ran in Halls A and B during the 6 GeV era, and  
consists of four inclusive doubly-polarized experiments: two in Hall A~\cite{Alcorn:2004sb}  
(E97-110 and E08-027) and two in Hall B~\cite{Mecking:2003zu} (E03-006  and E06-017, grouped under the 
EG4 denomination). Its main goal is to provide data to check spin-dependent calculations of 
Chiral Effective Field Theory ($\chi$EFT), an effective approach to QCD that should 
describe it at low energy/momentum, in particular at low $Q^2$ ($Q^2$ is the absolute value of the square of 
the 4-momentum transferred from  the beam to the target). 
This program is the continuation of a previous program at intermediate $Q^2$ which ran late in the late 1990s-early 2000s
and that had started to reach into the $\chi$EFT domain. This initial program consisted of two experiments,
E94-010 in Hall A, which measured  moments of the spin structure functions $g_1$ and $g_2$ on the neutron 
down to $Q^2 = 0.1$ GeV$^2$~\cite{Amarian:2002ar},  
and EG1 in Hall B, which measured of $g_1$  and its moments on proton and neutron down to $Q^2 = 0.05$ 
GeV$^2$~\cite{Yun:2002td}. The goal of the intermediate $Q^2$ program was to map the transition between the partonic to 
hadronic description of the strong force.
We will discuss these two programs. 
Other experiments on the nucleon spin structure at JLab are reviewed in~\cite{Deur:2018roz}

\section{Moments and spin sum rules}
The chief observables measured by the low- and intermediate-$Q^2$ spin programs are moments of $g_1$ and $g_2$ and
their associated sum rules. Of particular interest are the:
\vspace{-0.2cm}
\begin{itemize}
\item Gerasimov-Drell-Hearn sum rule (GDH)~\cite{Gerasimov:1965et}:
\vspace{-0.2cm}
\begin{equation}
\vspace{-0.2cm}
\int_{\nu_0}^{\infty}\frac{\sigma_{1/2}(\nu) - \sigma_{3/2}(\nu)}{\nu}d\nu=-\frac{4\pi^2 s \alpha\kappa^2}{M^2}, \nonumber
\end{equation}
where $ \sigma_{3/2}$ and $ \sigma_{1/2}$  denote the photoproduction cross sections for which the 
photon spin is parallel or antiparallel to the target spin $s$, respectively, 
$\nu$ is the photon energy, $\nu_0$ is the inelastic threshold, 
$M$ is the target mass, and $\alpha$ the QED coupling.  
The sum rule, derived for real photons, was later extended to $Q^2>0$~\cite{Anselmino:1988hn} by several type of:

\item Generalized GDH integrals, e.g.:
\begin{eqnarray}
\hspace{-0.8cm}
I_{TT}(Q^2) & = & \frac{M^2}{4\pi^2\alpha}\int_{\nu_0}^{\infty}\frac{\kappa_f}{\nu}\frac{\sigma_{1/2}(\nu,Q^2)-\sigma_{3/2}(\nu,Q^2)}{\nu}d\nu \nonumber
\\
& = & \frac{2M^2}{Q^2}\hspace{-0.15cm}\int_0^{x_0}\hspace{-0.1cm}\big[ g_1(x,Q^2)-\frac{4M^2}{Q^2}x^2g_2(x,Q^2)\big]dx   \nonumber
\label{eq:gdhsum_def1}
\end{eqnarray}
where  $\kappa_f$ the virtual photon flux~\cite{Deur:2018roz, Anselmino:1988hn} and $x=Q^2/(2M\nu)$ is the Bjorken scaling variable. 
The full sum rule was eventually extended~\cite{Ji:1999mr} as the:
\item Generalized GDH sum rule:
\vspace{-0.2cm}
\begin{equation}
\vspace{-0.2cm}
\Gamma_1(Q^2) \equiv \int_{0}^{x_0} \hspace{-0.1cm}g_1(x,Q^2) dx=\frac{Q^2 S_1}{8},  \nonumber
\end{equation}
with $S_1$ the polarized covariant VVCS amplitude~\cite{Ji:1999mr}. This sum rule is related to the:
\item Bjorken sum rule~\cite{Bjorken:1966jh}:
\vspace{-0.2cm}
 \begin{equation}
 \vspace{-0.1cm}
\Gamma_1^{p-n}(Q^2)=\frac{g_A}{6}\bigg[1-\frac{\alpha_{{\rm {s}}}(Q^2)}{\pi}-3.58\left(\frac{\alpha_{{\rm {s}}}(Q^2)}{\pi}\right)^2 
-20.21\left(\frac{\alpha_{{\rm {s}}}(Q^2)}{\pi}\right)^{3} +... \bigg]+O(1/Q^2), \nonumber
\label{eq:genBj}
\end{equation}
where $g_A$ is the nucleon axial charge, $\alpha_{\rm {s}} (Q^2)$ is the strong coupling~\cite{Deur:2016tte} that corrects
the sum rule for DGLAP evolution~\cite{Baikov:2010je}. $O(1/Q^2)$ are higher twists corrections.

These sum rules involves first moments. Others involving higher moments are the  

\item Generalized forward spin polarizability sum rule~\cite{GellMann:1954db}:
 \vspace{-0.2cm}
\begin{equation}
\vspace{-0.3cm}
\gamma_0(Q^2) =  \frac{16\alpha M^2}{Q^{6}}\int_0^{x_0}x^2\Bigl[g_1(x,Q^2)-\frac{4M^2}{Q^2}x^2g_2(x,Q^2)\Bigr]dx.  \nonumber
\label{eq:gamma_0}
\end{equation}
\item $LT$-polarizability sum rule:
 \vspace{-0.2cm}
 \begin{equation}
 \vspace{-0.2cm}
\delta_{LT}(Q^2)  =   \frac{16\alpha M^2}{Q^{6}}\int_0^{x_0}x^2\Bigl[g_1(x,Q^2)+g_2(x,Q^2)\Bigr]dx.
\label{eq:delta_{LT} SR}
\end{equation}
\item Burkhardt-Cottingham (BC) sum rule~\cite{Burkhardt:1970ti}:
 \vspace{-0.2cm}
 \begin{equation}
 \vspace{-0.2cm}
\Gamma_2(Q^2) \equiv  \int_0^1 g_2(x,Q^2)dx=0.
\label{eq:BC SR}
\end{equation}
\item $d_2$ sum rule:
~~~~~~~$d_2(Q^2)  =  \int_0^1 x^2 \big[ 2 g_1(x,Q^2)+3g_2(x,Q^2) \big] dx.$

\end{itemize}

Apart from the GDH sum rule which stands at $Q^2=0$, all these sum rules are valid for any $Q^2$. 
Thus, QCD can be studied by measuring a sum rule integral at various $Q^2$ 
and comparing it to the other sum rule side (``static" side) computed using techniques adapted to the $Q^2$ regime: 
at large  $Q^2$, pQCD and OPE; 
at intermediate $Q^2$, lattice QCD;
and at low $Q^2$, effective analytical approaches to non-perturbative QCD, such as $\chi$EFT.

\section{Lessons from the JLab intermediate $Q^2$ program}
In Hall A, E94-010 measured $\Gamma^n_1$, $\Gamma^n_2$, $\gamma^n_0$ and $\delta^n_{LT}$ 
in the $0.1 \leq Q^2 \leq 0.9$ GeV$^2$ range~\cite{Amarian:2002ar}. 
The Hall A neutron information was extracted from a $^3$He target, which is polarized using optical pumping and 
spin-exchange techniques. 
The lowest $Q^2$ data were compared to $\chi$EFT predictions~\cite{Bernard:1992nz, Ji:1999pd, Kao:2002cp}, 
but only $\gamma^n_0$ agreed with them.  ($\Gamma^n_2$ is not compared with $\chi$EFT, since the ``static side" of the 
BC sum rule, Eq.~(\ref{eq:BC SR}), is trivial). 
Chiefly surprising was the disagreement for $\delta_{LT}$, since it was expected to be reliably predicted by $\chi$EFT
due to the (supposed at the time) lack of $\Delta_{1232}$ resonance contribution, which was either not included in 
$\chi$EFT calculations, or included phenomenologically. Furthermore, the additional $x^2$ weighting in the 
$\delta_{LT}$ integral, Eq.~(\ref{eq:delta_{LT} SR}), reduces the usual experimental uncertainty due to the 
unmeasured low-$x$ part of the integral.
The discrepancy became known as the ``$\delta_{LT}$ puzzle".  
Data on $g^{^3He}_1$ and $g^{^3He}_2$ are also available.
%

In Hall B, EG1 provided  $\Gamma^p_1$, $\Gamma^n_1$, $\gamma^p_0$ and $\gamma^n_0$
in the $0.05 \leq Q^2 \leq 3$ GeV$^2$ domain~\cite{Yun:2002td}. 
The proton data were obtained using a longitudinally polarized NH$_3$ DNP target. A ND$_3$ target 
provided the neutron information. The lack of transverse polarization prevented 
to measure  $\Gamma_2$ and $\delta_{LT}$. Here again, only $\gamma^n_0$ agreed with the $\chi$EFT
predictions available at the time~\cite{Bernard:1992nz, Ji:1999pd, Kao:2002cp}.

Halls A and B data were combined to form the Bjorken sum $\Gamma_1^{p-n}$ and to provide an isospin analysis of the
data~\cite{Deur:2004ti}. The resulting $\Gamma_1^{p-n}$ agrees with 
$\chi$EFT~\cite{Bernard:1992nz, Ji:1999pd, Kao:2002cp}, validating
the argument that $\chi$EFT should reliably predict $\Gamma_1^{p-n}$  since 
the $\Delta_{1232}$  does not contribute to it~\cite{Burkert:2000qm}.

In all, the  conclusions that emerged from the first generation of experiments~\cite{Amarian:2002ar, Yun:2002td} and of 
$\chi$EFT predictions~\cite{Bernard:1992nz, Ji:1999pd, Kao:2002cp} were: 
\vspace{-0.2cm}
\begin{itemize}
\item The validity  domain of $\chi$EFT is smaller than the several tenths of GeV$^2$ initially hoped for, 
possibly only up to $\approx 0.1$ GeV$^2$ (this depends on the observable).
\vspace{-0.2cm}
\item There are no precise data below $\approx 0.1$ GeV$^2$: the experiments were designed 
for higher $Q^2$.
\vspace{-0.2cm}
\item The discrepancy for $\delta^n_{LT}$ is puzzling. There is no data for $\delta^p_{LT}$.
\end{itemize}
\vspace{-0.2cm}

\noindent The comparison between data and $\chi$EFT is summarized in Table~\ref{xpt-comp}. 

This state of affairs showed the necessity of an experimental program optimized to cover the chiral domain, and for
improved $\chi$EFT calculations.
\begin{table}[!htbp]
\vspace{-0.4cm}
{
\caption{Comparison between the first generation of moment data and of $\chi$EFT predictions. 
The bold fonts denote moments for which
$\chi$EFT was expected to provide robust predictions. ``{\color{blue} \bf{A}}" means that data and calculations agree up to at least
$Q^2=0.1$ GeV$^2$, ``{\color{red} \bf{X}}" that they disagree and ``-" that no calculation was available.
{\scriptsize \emph{p+n}} indicates either deuteron data without deuteron break-up contribution, 
or proton+neutron moments added together with neutron information either from D or $^3$He.
\label{xpt-comp}}
\vspace{0.1cm}
\begin{tabular}{|c|c|c|c|c|c|c|c|c|c|c|c|}
\hline 
Ref. & $\Gamma_1^p$ & $\Gamma_1^n$ & $\pmb{\Gamma_1^{p-n}}$ & $\Gamma_1^{p+n}$ &  $\gamma_0^p$ & $\gamma_0^n$ & $\pmb{\gamma_0^{p-n}}$ & $\gamma_0^{p+n}$ & $\pmb{\delta_{LT}^n}$ & $d_2^p$ & $d_2^n$ \tabularnewline
\hline
\hline 
Bernard {\emph{et al.}}~\cite{Bernard:1992nz} 
& {\color{red}\bf{X}} & {\color{red}\bf{X}} & {\color{blue}\bf{A}} & {\color{red}\bf{X}} & {\color{red}\bf{X}} & {\color{blue}\bf{A}} & {\color{red}\bf{X}} & {\color{red}\bf{X}}  & {\color{red}\bf{X}} & - & {\color{red}\bf{X}}\tabularnewline
\hline 
Ji {\emph{et al.}}~\cite{Ji:1999pd}  
& {\color{red}\bf{X}} & {\color{red}\bf{X}} & {\color{blue}\bf{A}} & {\color{red}\bf{X}} & - & - & - & - & - & - & - \tabularnewline
\hline 
Kao {\emph{et al.}}~\cite{Kao:2002cp}  
& - & - & - & - & {\color{red}\bf{X}} & {\color{blue}\bf{A}} & {\color{red}\bf{X}} & {\color{red}\bf{X}} & {\color{red}\bf{X}} & - & {\color{red}\bf{X}}\tabularnewline
\hline
\end{tabular}
\vspace{-0.5cm}
}
\end{table}
\section{The JLab low $Q^2$ program}
Experiments at low $Q^2$ were conducted at JLab to address the issues and puzzles just discussed.
E97-110~\cite{E97110} and E08-027~\cite{g2p} ran in Hall A and the EG4 run group ran in Hall B~\cite{Adhikari:2017wox}. 
\subsection{Experiment E97-110}
The main goal of E97-110~\cite{E97110} is to measure the generalized GDH sums for the neutron and $^3$He at $0.02 \leq Q^2 \leq 0.3$ 
GeV$^2$. The experiment ran in JLab's Hall A and data were taken at two scattering angles, $6^\circ$ and $9^\circ$, 
using a polarized ($\approx 85\%$) electrons of energies  4.2, 2.8, 2.2, 2.1, 1.5 and 1.2 GeV for the $6^\circ$ data and 
4.4, 3.8, 3.3, 2.2, and 1.2 GeV for the $9^\circ$ data. The $^3$He target can be polarized longitudinally or transversally 
(in the horizontal plane) with respect to the beam, which allows to measure $g_1$ and $g_2$. 
The time shared between longitudinal and transverse data taking was optimized for 
maximal precision on the GDH integrant $\sigma_{TT} \propto g_1 - \frac{Q^2}{\nu^2} g_2$.
To reach low $Q^2$ while covering enough $x$ range to form integrals, 
small scattering angles are required. They were reached by adding a ``septum" magnet  to one of the Hall A 
spectrometers. This lowered its minimal angle from $12.5^\circ$  to $6^\circ$.
$g_1$ and $g_2$ were extracted using the difference of  polarized cross-sections:
\vspace{-0.2cm}
\begin{equation}
\vspace{-0.1cm}
\sigma^{\downarrow\Uparrow} - \sigma^{\uparrow\Uparrow}=\frac{4\alpha^2}{MQ^2}\frac{E'}{E\nu}\left[g_1(E+E'\cos\theta)-Q^2\frac{g_2}{\nu}\right],
~~~~
\sigma^{\downarrow\Rightarrow} - \sigma^{\uparrow\Rightarrow} =\frac{4\alpha^2}{MQ^2}\frac{E'^2}{E\nu}\sin\theta\left[g_1+2E\frac{g_2}{\nu}\right],
\nonumber
\end{equation}
which advantageously cancels contributions from unpolarized materials in the target or the beamline.
The $\downarrow$ and $\uparrow$ represent the beam helicity
while $\Downarrow$ , $\Uparrow$ and $\Rightarrow$  indicate the direction of the target polarization.
E97-110 is described in more details in N. Ton's contribution to these proceedings.

\begin{figure}[ht]
\centering
\vspace{-0.2cm}
\includegraphics[width=13.0cm]{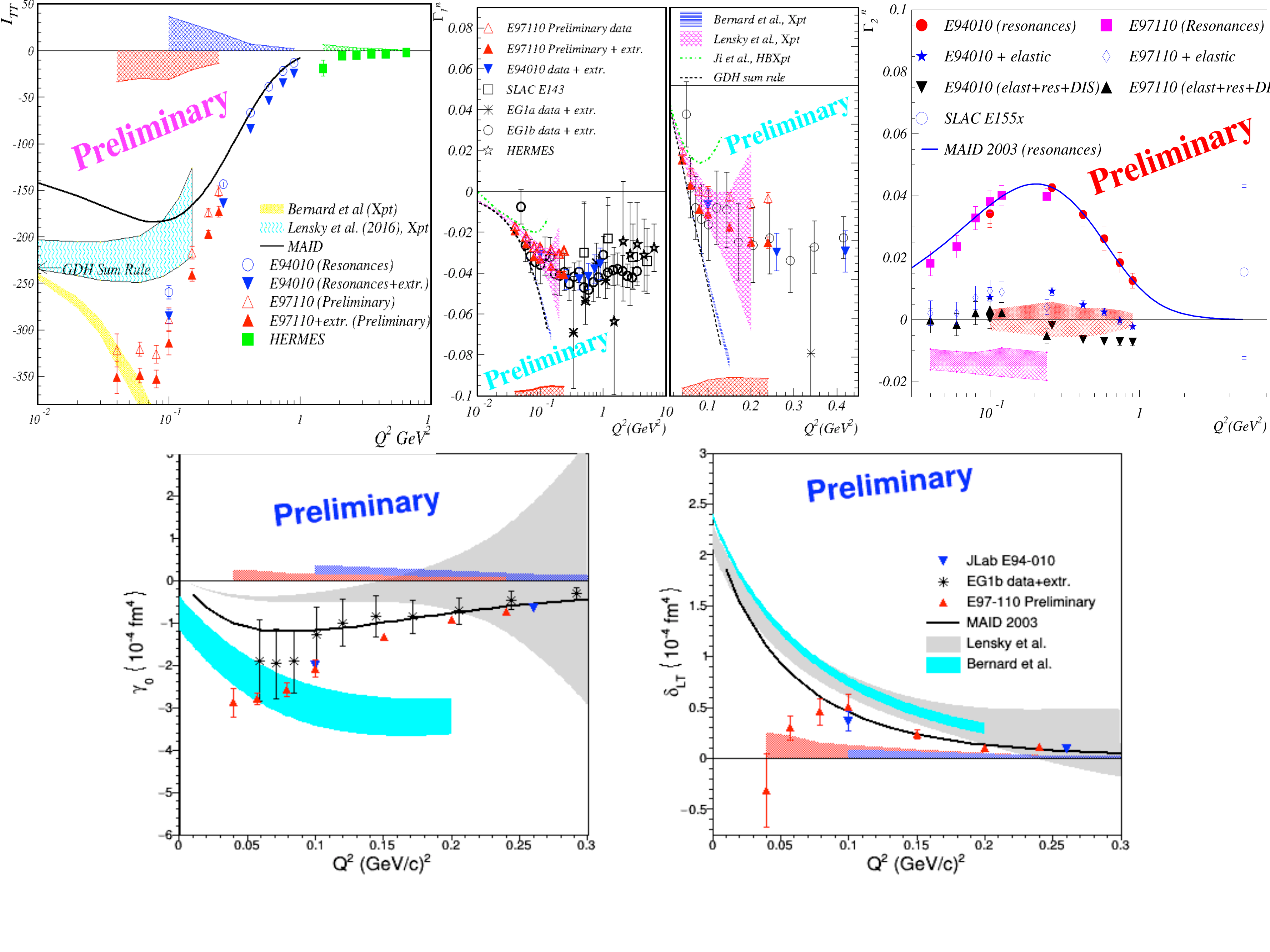}
\vspace{-0.5cm}
\caption{\label{fig:E97-110}
Preliminary neutron results from E97-110 for $I_{TT}^n$ (top left), $\Gamma_1^n$ (top center), 
$\Gamma_2^n$ (top right), $\gamma_0^n$ (bottom left) and $\delta_{LT}^n$ (bottom right). 
The open symbols are for the measured part of the moments. The solid ones include an estimate of the
unmeasured low-$x$ contribution (and elastic for $\Gamma_2^n$). The low-$x$ contribution is
negligible for $\gamma_0^n$ and $\delta_{LT}^n$. Also shown are
recent $\chi$EFT calculations, the 
MAID model and results of earlier experiments E155, HERMES and E94-010 at larger $Q^2$. }
\vspace{-0.5cm}
\end{figure} 

Fig.~\ref{fig:E97-110} shows preliminary results for $I_{TT}^n$, $\Gamma_1^n$, $\Gamma_2^n$, and
higher moments $\gamma_0^n$ and $\delta_{LT}^n$. E97-110 agrees well with the E94-010 and EG1b data (when available). 
The $I_{TT}^n$ data disagree with the $\chi$EFT result of Lensky {\emph{et al.}}~\cite{Lensky:2014dda} 
and agree with that of Bernard {\emph{et al.}}~\cite{Bernard:2012hb} for the lowest $Q^2$ points. 
If the GDH sum rule holds then a sharp turn-over must occur below $Q^2 \approx 0.05$ GeV$^2$.
Compared to E94-010, the lowest $Q^2$ value has been reduced by factor of $\approx 2.5$,
which provides data to test of $\chi$EFT well into the chiral domain.
More data down to $Q^2= 0.02$ GeV$^2$ are being analyzed and will check further  
$\chi$EFT and the status of the turn-over.
The $\Gamma_1^n$ data agree with Lensky {\emph{et al.}} over a large $Q^2$-range
and agree with Bernard {\emph{et al.}} for a smaller range. 
$\Gamma_2^n$ data seem to agree with the BC sum rule. However, 
the unmeasured low-$x$ part, difficult to assess, causes a large uncertainty.
The bottom plots in Fig.~\ref{fig:E97-110} show higher moments.  As for $I_{TT}^n$, the $\gamma_0^n$ 
data disagree with Lensky {\emph{et al.}} and agree with Bernard {\emph{et al.}} for the lowest $Q^2$ points. 
For $\delta_{LT}^n$, while the new  $\chi$EFT calculations agree with E94-010 and thus seem to have resolved 
the $\delta_{LT}^n$ puzzle, E97-110 data at lower $Q^2$ may renew the puzzle. 

\subsection{Experiment group EG4}

EG4~\cite{Adhikari:2017wox} consists of two experiments, E03-006 (proton) and E06-017 (neutron),
whose goal is to measure the generalized GDH sum at very low $Q^2$. they ran in Hall B
using polarized electrons of energies of 3.0, 2.3, 2.0, 1.3 or 1.0 GeV scattering off a longitudinally 
polarized target containing either NH$_3$ or ND$_3$. The H or D were polarized using the DNP technique.
This allowed to measure $g_1^p$ and $g_1^D$, from which neutron information will be extracted.
As mentioned, measuring moments at low $Q^2$ asks for high beam energy and small scattering angle.
The latter was reached by setting the CLAS field polarity to outbend electrons
and by installing the target  $1$ m upstream its nominal location.
$g_1$ is extracted using cross-section difference. This demands a well controlled (i.e high)
detection efficiency at small angles. For this purpose, a new Cerenkov counter 
was installed in one CLAS sector. It allowed to measure cross-sections down to $6^\circ$.
EG4 is presented in more details in K. Slifer's contribution to these proceedings.

\begin{figure}[ht]
\centering
\vspace{-0.2cm}
\includegraphics[width=12.0cm]{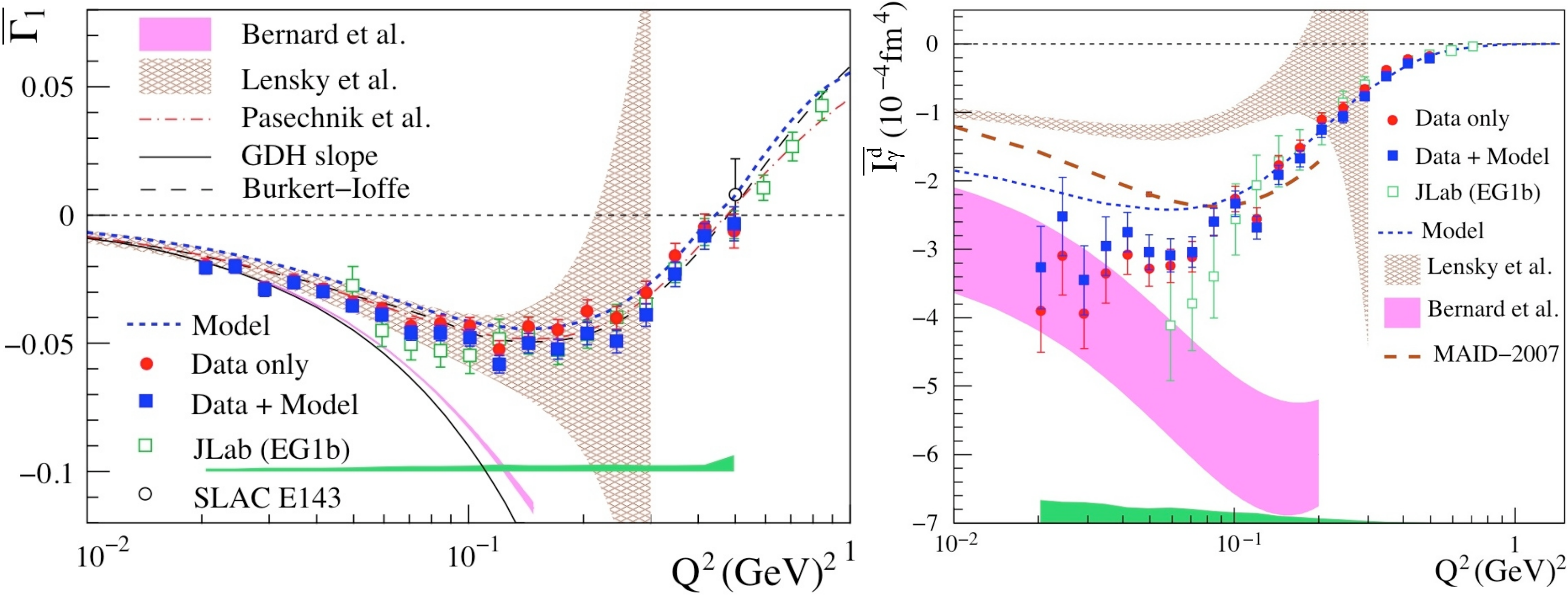}
\vspace{-0.3cm}
\caption{\label{fig:EG4}
EG4 results for $\overline{\Gamma_1^d}$ (left) and $\overline{\gamma_0^d}$ (right).
The circles are for the measured part of the integral, while the squares include an estimate of the
unmeasured low-$x$ contribution. Also shown are $\chi$EFT 
calculations, several phenomenological models and earlier results from E143 and EG1b at larger $Q^2$.}
\vspace{-0.2cm}
\end{figure} 

Fig.~\ref{fig:EG4} shows results for $\overline{\Gamma_1^d}$ and $\overline{\gamma_0^d}$.
The bar means that the deuteron photodisintegration contribution is not included in the moments. Hence they
represent approximately the sum of the proton and neutron moments.
The EG4 and EG1b data agree well. 
The $\overline{\Gamma_1^d}$ data agree with the $\chi$EFT results of Lensky {\emph{et al.}} and also that of
Bernard {\emph{et al.}} but only for the lowest $Q^2$ points.
The models of Pasechnik {\emph{et al.}}~\cite{Pasechnik:2010fg} and Burkert-Ioffe~\cite{Burkert:1992tg} agree well with the data.
The $\overline{\gamma_0^d}$ data disagree with the Lensky {\emph{et al.}} results. The ones from
Bernard {\emph{et al.}} again agree for lowest $Q^2$ points. The Maid model~\cite{Drechsel:1998hk} 
disagrees below $Q^2<0.1$ GeV$^2$.
The low $Q^2$ reach of the data is decreased by a factor of $\approx$2.5, compared to the earlier experiments,
testing $\chi$EFT well into the chiral domain. The EG4 data also display a much improved precision.
Analysis for $\overline{\Gamma_1^p}$, and $\overline{\gamma_0^p}$ for the proton is on-going, with final results expected within 2019.

\subsection{Experiment E08-027}

The main goal of E08-027~\cite{g2p} is to measure  $\delta^p_{LT}$  down to $Q^2$=0.01 GeV$^2$.
$\Gamma_1^p$, $\Gamma_2^p$, $\gamma_0^p$, $g_1^p$ and $g_2^p$ are also being extracted from the data. 
Beside testing for the first time $\chi$EFT using $\delta_{LT}^p$, E08-027 provides the first high-precision data to 
study the BC sum rule on the proton and the first $g_2^p$ data at low enough $Q^2$ to be useful to  
proton hyperfine studies. The experiment ran in Hall A using  polarized electrons of 3.4, 2,3, 1.7 or 
1.2  GeV. A NH$_3$ target similar to that of EG4 was used, but with transverse polarization capability, in addition to the 
longitudinal one, to access $g_2^p$ and consequently $\delta_{LT}^p$. It was the first use of such 
target in Hall A and new equipment was needed to characterize the beam of low current imposed by the target.
Chicanes and a local beam dump were installed to accommodate the target transverse magnetic field. The
small angle necessary to reach  low $Q^2$ was provided by septum magnets.
$g_1^p$ and $g_2^p$ are obtained from cross-section differences.
E08-027 is discussed more extensively in K. Slifer's contribution to these proceedings.

The preliminary $\delta_{LT}^p$ data, ranging $0.045 \leq Q^2 \leq 0.13$ GeV$^2$, agree well with 
the Lensky {\emph{et al.}}  calculations but not with the Bernard {\emph{et al.}} ones. The data also agree
with the MAID model~\cite{Drechsel:1998hk}. 
$\gamma_0^p$ is available at $Q^2=0.045$ GeV$^2$ and show tensions with the EG1b data
(which agree with Lensky {\emph{et al.}}) and Bernard {\emph{et al.}}
(which disagrees with EG1b and Lensky {\emph{et al.}}).

\section{Current state of testing $\chi$EFT with spin sum rules.}

The experiments above are testing $\chi$EFT well into the chiral domain and with improved precision. On the theory side,
two recent predictions are available (Lensky {\emph{et al.}}~\cite{Lensky:2014dda} 
and Bernard {\emph{et al.}}~\cite{Bernard:2012hb}). Table~\ref{xpt-comp2} summarizes how they compare for
$Q^2 \leq 0.1$ GeV$^2$. It shows that a satisfactory description of nucleon spin structure  by  $\chi$EFT
remains challenging. Some observables, such as  $\Gamma_1^{p-n}$, are well described over 
large ranges, while others, such as $\delta_{LT}^n$, remain refractory to a $\chi$EFT description. Others have mixed success,
agreeing with one $\chi$EFT calculation but not the other. The two calculations generally disagree with each other. 
Since several of the experimental results discussed are preliminary, one has wait for the final results to  confirm the above
conclusion.
\begin{table}
\vspace{-0.5cm}
{
\caption{Same as Table~\ref{xpt-comp} but for the newest experiments and $\chi$EFT results. The $^*$ signals
preliminary data.
\label{xpt-comp2}}
\vspace{0.2cm}
\begin{tabular}{|c|c|c|c|c|c|c|c|c|c|c|c|c|}
\hline 
Ref. & $\Gamma_1^p$ & $\Gamma_1^n$ & $\pmb{\Gamma_1^{p-n}}$ & $\Gamma_1^{p+n}$ &  $\gamma_0^p$ & $\gamma_0^n$ & $\pmb{\gamma_0^{p-n}}$ & $\gamma_0^{p+n}$ & $\pmb{\delta_{LT}^p}$ & $\pmb{\delta_{LT}^n}$ & $d_2^p$ & $d_2^n$ \tabularnewline
\hline
\hline 
Bernard {\emph{et al.}}~\cite{Bernard:2012hb}  
& {\color{red}\bf{X}} & {\color{red}\bf{X}} & {\color{blue}\bf{A}} & {\color{red}\bf{X}} & {\color{red}\bf{X}} & {\color{blue}\bf{A}} & {\color{red}\bf{X}} & {\color{red}\bf{X}} & {\color{red}\bf{X*}} & {\color{red}\bf{X*}} & - & -\tabularnewline
\hline 
Lensky {\emph{et al.}}~\cite{Lensky:2014dda} 
& {\color{red}\bf{X}} & {\color{blue}\bf{A}} &  {\color{blue}\bf{A}} & {\color{blue}\bf{A}} &  {\color{blue}\bf{A}} & {\color{red}\bf{X}} & {\color{red}\bf{X}} & {\color{red}\bf{X}} & {\color{blue}\bf{A*}} & {\color{red}\bf{X*}} & NA &  {\color{blue}\bf{A}}\tabularnewline
\hline
\end{tabular}
\vspace{-0.5cm}
}
\end{table}

\section{Summary and perspective}
The JLab low $Q^2$ experimental program was the last stage to complete the experimental 
mapping from high to low $Q^2$ of $g_1$, $g_2$ and their moments for nucleons and light nuclei. It 
complements the intermediate $Q^2$ program of JLab and the high $Q^2$ programs of SLAC, CERN and DESY.

The EG4 deuteron data are published~\cite{Adhikari:2017wox}. The others data, from E97-110, EG4-proton and 
E08-027 are in final analysis stage, with preliminary results available and final results expected within 2019. 
Thus, a comprehensive set of data (both nucleons, and both $g_1$ and $g_2$ moments and their 
combinations)  will be available shortly to test $\chi$EFT. Meanwhile, theory groups are improving
calculations and studying the origin of the difference between their predictions. 
A preliminary conclusion is that in spite of notable improvements compared to the early calculations, 
 $\chi$EFT describes the nucleon spin structure at low $Q^2$ with mixed success, 
depending on the specific theoretical approach and on the observable. 
The data analyses must be finalized before to draw 
firm conclusions  but it seems that describing the nucleon spin structure remains a challenge for $\chi$EFT.

\end{document}